\documentclass[10pt,reqno]{amsart}
\usepackage{amssymb,amsthm,amsmath}
\usepackage{graphicx}
\usepackage{xcolor}
\usepackage{hyperref}
\usepackage{indentfirst}
\usepackage{csquotes}
\usepackage{geometry}
\usepackage{setspace}
\usepackage{caption}
\usepackage{cite}
\usepackage{tcolorbox}
\usepackage{tikz}
\usepackage{float}
\usepackage{multirow}
\usepackage{rotating}
\usepackage{adjustbox}
\usepackage{lscape}
\usepackage{mathptmx}
\usepackage{booktabs}
\usepackage{amsaddr}
\usepackage{pdflscape}
\usepackage{float}

\hypersetup{
	colorlinks=true,
	linkcolor=black,
	filecolor=black,
	urlcolor=black
}

\begin{document}
	\makeatletter
	\let\mathindent\@centering
	\makeatother
	
	\title{Non-Commutative Phase-Space Effects in Fermionic String Theory} %%%%%%%%%%%%
	\author[Initial Surname]{Mohamed Adib Abdelmoumene}
	\date{\today}
	\address{Physics Department, Laboratoire de Physique Mathematique et Physique Subatomique, LPMPS, University Mentouri Constantine 1, Constantine, Algeria}
	\email{abdelmoumene.medadib@umc.edu.dz}
	\author[Initial Surname]{Nadir Belaloui}
	\date{\today}
	\address{Physics Department, Laboratoire de Physique Mathematique et Physique Subatomique, LPMPS, University Mentouri Constantine 1, Constantine, Algeria}
	\email{belaloui.nadir@umc.edu.dz}

	\maketitle
\begin{center}
	\textbf{Accepted for publication in Acta Physica Polonica B.}
\end{center}
	
	\begin{abstract}
		We study free open fermionic strings on a non-commutative phase space. Modified super-Virasoro algebras in both Ramond and Neveu-Schwarz sectors acquire non-commutativity anomalies, and this noncommutativity also breaks Lorentz symmetry and give a non-diagonal mass operator. Redefining the Fock space diagonalizes the mass operator. Extra constraints on non-commutativity parameters cancel the anomalies, restore the standard spectrum and make the GSO projection possible.\\
	\end{abstract} %%%%%%%%%
\section{Introduction}\label{sec1}

The non-commutativity was first studied by Connes \cite{bib1}, which was considered as a relation between many connections in physics, and then in string theory \cite{witten,shahin1}. \\
One can also obtain non commutativity when we can consider a string interacting with an antisymmetric B-field or a N-S B-field \cite{shahin2,shahin3, witten1,bib2,bib3,bib4,bib5,RevModPhys.73.977}. The main results are the fact that the equations of motion do not depend on the B-field while the Noether currents, the boundaries conditions and in particular, the momentum do. A great number of works \cite{shahin2,shahin3, witten1,bib2,bib3,bib4,bib5,RevModPhys.73.977,bib6,bib7} have investigated the covariant quantization of this theory and shows that the coordinates extremities of the string are non-commutative and that the physical states are subject to the Virasoro conditions, which depend on the B-field.
{Throughout these works, the background antisymmetric field $B_{\mu\nu}$ is understood to have support only along the Neumann directions of the $D_p$-brane, $B_{ij}$ with $i,j = 0,\ldots,p$ labeling directions along the brane worldvolume, since only these components contribute to the open-string boundary variation and modify the effective worldvolume dynamics while mixed or Dirichlet components do not .
	As a result, the effect of the $B$-field is fully captured by an effective field strength $F_{ij}$, and the induced non-commutative structure arises solely on the D-brane worldvolume.\\}
The B-field dependence of the momentum makes difficult the definition of the light cone gauge. This difficulty is resolved for the closed bosonic string \cite{bib8} when it wraps a compactified dimension on which we have the only non-zero component of the B-field. The periodicity conditions allow then the use of the light cone gauge.\\
Another way consist to consider a string propagating in a non-commutative worldsheet, Our motivation for investigating noncommutativity on the worldsheet stems from the fact that when the latter propagates in a noncommutative spacetime, it inherits this noncommutative structure and consequently becomes a two-dimensional noncommutative space \cite{Kamani_2002, KAMANI_2004, Kamanii_2002}. The commutation relations between the modes and the Virasoro algebra are modified. These modifications will affect the mass spectrum and the Lorentz invariance.\\
It is important to emphasize that such a theory is inherently non-conformal and lacks Lorentz invariance. Nevertheless, it is worthwhile to investigate this framework in order to explore possible ways of addressing these issues by taking advantage of the existence of two independent noncommutativity parameters $\theta_{(m)}$ and $\gamma_{(m)}$ and examining potential relations between them. This is the principal  motivation for introducing noncommutativity in the full phase space, rather than only between coordinate variables, to explore whether the Virasoro (and Lorentz) algebras can be restored or not, by imposing a consistent relation between the two deformation parameters, $\theta_{(m)}$ and $\gamma_{(m)}$. Our results show that the Virasoro algebra can indeed be recovered in a noncommutative phase space when the deformation is restricted to the center-of-mass variables. On the other hand, in order to restore the Lorentz algebra completely, one must return to an ordinary (commutative) phase space.\\

{In contrast with the Seiberg–Witten framework \cite{witten1}, where noncommutativity arises in the target-space coordinates of open strings propagating in a background antisymmetric $B_{\mu\nu}$ field, the deformation considered here is canonical rather than geometric: it affects the equal-time commutation relations of the string coordinates and their conjugate momenta, while leaving the target-space geometry itself commutative. As a result, the theory does not define a noncommutative quantum field theory on spacetime, and the well-known consistency and regularization issues discussed in noncommutative gauge theories—such as $UV/IR$ mixing, unitarity violation, or renormalization ambiguities—do not arise within the present framework. Moreover, since the super-Virasoro generators depend explicitly on both coordinate and momentum variables, a deformation restricted solely to the coordinate sector would generically obstruct the closure of the algebra. A simultaneous deformation of coordinates and momenta therefore seems to be the minimal extension required to preserve the algebraic consistency perturbatively.}\\

In this paper, we investigate a fermionic open string theory that moves in a non-commutative phase-space. In section 2, we describe how noncommutativity in the phase-space is obtained from noncommutativity in the worldsheet. In section 3, the commutation relations for the string coordinates are postulated and the oscillator algebra are obtained. In section 4, we calculate the modified super-Virasoro algebra for both of the Ramond and Neveu-Schwarz sectors. In section 5, we deduce the modified Lorentz algebra. In section 6, mass spectrum and the GSO projection are discussed. Finally we summary our results.\\

\section{Non-commutative phase-space from a non-commutative worldsheet}\label{sec2}
In the standard formulation of string theory, the worldsheet is a two-dimensional manifold parameterized by the coordinates $\sigma^a = (\tau, \sigma)$. However, in a non-commutative framework, these coordinates satisfy the fundamental commutation relation \cite{KAMANI_2004,Kamanii_2002} giving by:
\begin{equation}
	[\sigma^a, \sigma^b] = i\theta^{ab}, 
\end{equation}
where $\theta^{ab}$ is an antisymmetric constant tensor representing the non-commutativity of the worldsheet. As a result, a deformation of the usual product of functions, introducing the Moyal star product\cite{KAMANI_2004,Kamanii_2002}:
\begin{equation}
	(f * g)(\sigma,\tau) = 
	f(\sigma,\tau)\, 
	e^{2 i \theta^{ab} \overleftarrow{\partial_a} \overrightarrow{\partial_b}}\, 
	g(\sigma,\tau)
\end{equation}
Expanding the exponential, we find:
\begin{equation}
	\begin{array}{l}
		f(\sigma ,\tau ) * g(\sigma ,\tau ) = f(\sigma ,\tau )g(\sigma ,\tau ) + \frac{i}{2}{\theta ^{ab}}{\partial _a}f{\partial _b}g
		- \frac{1}{8}{\theta ^{ab}}{\theta ^{cd}}\left( {{\partial _a}{\partial _c}f} \right)\left( {{\partial _b}{\partial _d}g} \right) + {\mathcal O}({\theta ^3})
	\end{array}
\end{equation}
For all calculations in the manuscript, we restrict ourselves to linear order in $\theta$, as higher orders produce higher-derivative (nonlocal) corrections which are beyond the present semiclassical treatment. This truncation is consistent with standard treatments of noncommutative field theory \cite{KAMANI_2004}.\\
The action of the superstring in this framework is modified as well:
\begin{equation}
	\label{s*}	S_{\ast} = - \frac{1}{4\pi\alpha'} \int d^2\sigma \left( \partial_a X^\mu \ast \partial^a X_\mu - i \bar{\psi}^\mu \ast \rho^a \partial_a \psi_\mu \right)	
\end{equation}
where the star product replaces the usual multiplication in all field interactions, and
\begin{equation}
	{\rho ^0} = \left( {\begin{array}{*{20}{c}}
			0&{ - i}\\
			i&0
	\end{array}} \right),
\end{equation}
\begin{equation}
	{\rho ^1} = \left( {\begin{array}{*{20}{c}}
			0&{  i}\\
			i&0
	\end{array}} \right)
\end{equation}
that verify the Clifford algebra given by: 
\begin{equation}
	\left\{ {{\rho ^\alpha },{\rho ^\beta }} \right\} =  - 2{\eta ^{\alpha \beta }}
\end{equation}
We will show how non-commutative structure propagates to the spacetime coordinates.\\

The non-commutativity of the worldsheet directly induces non-commutativity in the spacetime coordinates \cite{KAMANI_2004}. By using the star product, the commutator between spacetime coordinates takes this form:
\begin{equation}
	\begin{array}{l}
		{[{X^\mu }(\sigma ,\tau ),{X^\nu }(\sigma ',\tau )]_ * } = \theta {(2\alpha ')^{3/2}}\sum\limits_{n \ne 0} {\left[ \begin{array}{l}
				({p^\mu }\alpha _n^\nu  - {p^\nu }\alpha _n^\mu )\\
				{e^{ - in\tau }}\sin (n\sigma )
			\end{array} \right]} \\
		- 2\alpha '\sum\limits_{m \ne 0} {\sum\limits_{n \ne 0} {\left( \begin{array}{l}
					\frac{1}{{mn}}(\alpha _n^\mu \alpha _m^\nu  - \alpha _n^\nu \alpha _m^\mu ){e^{ - i(m + n)\tau }}
					\cos \left( {m(\sigma  + \frac{1}{2}n\theta )} \right)
					\cos \left( {n(\sigma ' - \frac{1}{2}m\theta )} \right)
				\end{array} \right)} } 
	\end{array}
\end{equation}
This result shows that the non-commutativity of the worldsheet introduces a deformed algebra for spacetime coordinates.\\

Since the momentum density is defined as:
\begin{equation}
	P^\mu(\sigma, \tau) = \frac{1}{2\pi\alpha'} \dot{X}^\mu
\end{equation}
and inherits again the non-commutative structure as $X^\mu$ , the corresponding commutator in the presence of a worldsheet non-commutativity is found to be:
\begin{equation}
	\begin{array}{l}
		{[{P^\mu }(\sigma ,\tau ),{P^\nu }(\sigma ',\tau )]_*} = 
		{\rm{         }}\frac{1}{{2\alpha '{\pi ^2}}}\sum\limits_{m \ne 0} {\sum\limits_{n \ne 0} {\left( {\begin{array}{*{20}{l}}
						{(\alpha _n^\mu \alpha _m^\nu  - \alpha _n^\nu \alpha _m^\mu ){e^{ - i(m + n)\tau }}}
						{\cos \left( {m(\sigma  + \frac{1}{2}n\theta )} \right)}
						{\cos \left( {n(\sigma ' - \frac{1}{2}m\theta )} \right)}
				\end{array}} \right)} } 
	\end{array}
\end{equation}
Thus, both of position and momentum variables obey the non-commutative relations. This will leads to a fundamental deformation of the phase-space, where both of position and momentum no longer satisfies the usual Poisson structure but instead obey a deformed algebra.\\
\section{Non-commutative phase-space and oscillator algebra} \label{SEC3}
Let us now consider a non-commutative phase-space described by the following non-commutative commutation relations \cite{bib10, bib11} :
\begin{equation}
	\label{1}\begin{array}{l}
		\left[ {{X^\mu }\left( {\tau ,\sigma } \right),{P^\nu }\left( {\tau ,\sigma '} \right)} \right] = i{\eta ^{\mu \nu }}\delta \left( {\sigma  - \sigma '} \right)\\
		\left[ {{X^\mu }\left( {\tau ,\sigma } \right),{X^\nu }\left( {\tau ,\sigma '} \right)} \right] = i{\theta ^{\mu \nu }} \left( {\sigma  - \sigma '} \right)\\
		\left[ {{P^\mu }\left( {\tau ,\sigma } \right),{P^\nu }\left( {\tau ,\sigma '} \right)} \right] = i{\gamma ^{\mu \nu }} \left( {\sigma  - \sigma '} \right)\\
		\left\{ {{\psi ^\mu }\left( {\tau ,\sigma } \right),{\psi ^\nu }\left( {\tau ,\sigma '} \right)} \right\} = {\eta ^{\mu \nu }}\delta \left( {\sigma  - \sigma '} \right)
	\end{array}
\end{equation}
Where ${P^\mu }\left( {\tau ,\sigma } \right) = \frac{1}{{2\pi \alpha '}}{\partial _\tau }{X^\mu }\left( {\tau ,\sigma } \right)$, ${\theta ^{\mu \nu }}$ represent the non-commutativity parameters of the space part and $\gamma ^{\mu\nu}$ the ones of the momentum part of the phase-space.\\

For the sake of simplicity, this noncommutativity is only postulated by assuming the existence of a noncommutative worldsheet as the origin of \eqref{1}, without worrying about the latter. Consequently, since the equation of motion remain unchanged (See Annexe \ref{appb}), the solutions and the Virasoro operators also remain unaffected.\\
The equations of motions are:
\begin{equation}
	\left( {\partial _\tau ^2 - \partial _\sigma ^2} \right){X^\mu } = {\partial _ + }\psi _ - ^\mu  = {\partial _ - }\psi _ + ^\mu  = 0
\end{equation}
where ${\partial _ \pm } = \frac{1}{2}\left( {{\partial _\tau } \pm {\partial _\sigma }} \right)$, while $\psi _ -$ and $\psi _ +$ are the right moving and the left moving components of $\psi^\mu$.\\
One can write the Fourier expansions for the variables  ${\theta ^{\mu \nu }}(\sigma - \sigma\prime)$ , $\gamma ^{\mu\nu}(\sigma - \sigma\prime)$ \cite{bib10}  and $X^{\mu}(\tau , \sigma)$, $\psi ^\mu(\tau ,\sigma)$ \cite{bib15,bib14,bib16} :
\begin{equation}
	\label{2}{\theta ^{\mu \nu }}\left( {\sigma  - \sigma '} \right) = \sum\limits_{n =  - \infty }^{ + \infty } {\theta _n^{\mu \nu }} {e^{in\left( {\sigma  - \sigma '} \right)}}
\end{equation}
\begin{equation}
	\label{3}{\gamma ^{\mu \nu }}\left( {\sigma  - \sigma '} \right) = \sum\limits_{n =  - \infty }^{ + \infty } {\gamma _n^{\mu \nu }} {e^{in\left( {\sigma  - \sigma '} \right)}}
\end{equation}
\begin{equation}
	\label{4}{X^\mu }\left( {\tau ,\sigma } \right) = {x^\mu } + 2\alpha '{p^\mu }\tau  + i\sqrt {2\alpha '} \sum\limits_{n \ne 0} {\frac{1}{n}\alpha _n^\mu \cos (n\sigma )} {e^{ - in\tau }}
\end{equation}
\begin{equation}
	\label{ff} \begin{array}{l}
		NS - \sec tor:{\rm{ }}{\psi ^\mu(\tau , \sigma) } = \frac{1}{{\sqrt 2 }}\sum\limits_{r \in Z + \frac{1}{2}} {b_r^\mu {e^{ - ir\left( {\tau  - \sigma } \right)}}} \\
		R - \sec tor:{\rm{ }}{\psi ^\mu(\tau , \sigma) } = \frac{1}{{\sqrt 2 }}\sum\limits_{n \in Z} {d_n^\mu {e^{ - in\left( {\tau  - \sigma } \right)}}} 
	\end{array}
\end{equation}
By the use of (\ref{2}), (\ref{3}), (\ref{4}) and (\ref{ff}), we can verify that the equations (\ref{1}) are equivalent to the following commutation relations of the oscillator algebra \cite{bib10} :
\begin{equation}
	\label{66}	
	\begin{array}{*{20}{l}}
		{\left[ {{p^\mu },{p^\nu }} \right] = i{\pi ^2}{\gamma _0}^{\mu \nu }}\\
		{\left[ {{x^\mu },{p^\nu }} \right] = i{\eta ^{\mu \nu }} - 2i{\pi ^2}\alpha '\tau {\gamma _0}^{\mu \nu }}\\
		{\left[ {{x^\mu },{x^\nu }} \right] = i{\theta _0}^{\mu \nu } - 4i{\pi ^2}{{\alpha '}^2}{\tau ^2}{\gamma _0}^{\mu \nu }}
	\end{array}
\end{equation}
\begin{equation}
	\label{6} \left[ {\alpha _m^\mu ,\alpha _n^\nu } \right] = \left( {m{\eta ^{\mu \nu }} + i \frac{{{(2\pi \alpha ')^2}}}{{2\alpha '}}{\gamma _n}^{\mu \nu } + i \frac{{{n^2}}}{{2\alpha '}}{\theta _n}^{\mu \nu }} \right){\delta _{n + m,0}}
\end{equation}
\begin{equation}
	\label{669} \left\{ \begin{array}{l}
		\left\{ {d_m^\mu ,d_n^\nu } \right\} = {\eta ^{\mu \nu }}{\delta _{m + n,0}}\\
		\left\{ {b_r^\mu ,b_s^\nu } \right\} = {\eta ^{\mu \nu }}{\delta _{r + s,0}}
	\end{array} \right.
\end{equation}

\section{Modified Super-Virasoro Algebra}\label{sec3}
We define the Virasoro generators in a quantized system as \cite{bib15,bib14,bib16,bib9}:\\
For Ramond sector :
\begin{equation}
	\label{11}	{L_m} = L_m^\alpha  + L_m^d = \left\{ \begin{array}{l}
		L_m^\alpha  = \frac{1}{2}\sum\limits_{n \in Z} {:{\alpha _{ - n}}{\alpha _{m + n}}:} \\
		L_m^d = \frac{1}{2}\sum\limits_{n \in Z} {\left( {n + \frac{1}{2}m} \right):{d_{ - n}}{d_{m + n}}:} 
	\end{array} \right.
\end{equation}
\begin{equation}
	\label{12}	{F_m} = \sum\limits_{n \in Z} {{\alpha _{ - n}}{d_{m + n}}} 
\end{equation}
Which represent the fermionic sector.\\
For Neveu-Schwarz sector :
\begin{equation}
	\label{13}	{L_m} = L_m^\alpha  + L_m^b = \left\{ \begin{array}{l}
		L_m^\alpha  = \frac{1}{2}\sum\limits_{n \in Z} {:{\alpha _{ - n}}{\alpha _{m + n}}:} \\
		L_m^b = \frac{1}{2}\sum\limits_{r \in Z+\frac{1}{2}} {\left( {r + \frac{1}{2}m} \right):{b_{ - r}}{b_{m + r}}:} 
	\end{array} \right.
\end{equation}
\begin{equation}
	\label{14}	{G_r} = \sum\limits_{n \in Z} {{\alpha _{ - n}}{b_{r + n}}} 
\end{equation}
Which represent the bosonic sector.\\
Because of the modifications in the oscillator algebra ({\ref{6}}), one can deduce the modified super-Virasoro algebras for both sectors,
\begin{equation}
	\begin{array}{l}
		[L_m^{(\alpha )},L_n^{(\alpha )}] = (m - n)L_{n + m}^{(\alpha )} +\frac{d}{{12}}m\left( {{m^2} - 1} \right){\delta _{m + n,0}} + R_{mn}
	\end{array}
\end{equation}
{$n$ represents the integer modes of the bosonic oscillators, where $n \in \mathbb{Z}$}, and $R_{mn}$ represent the anomaly part due to the non-commutativity, defined by:
\begin{equation}
	\label{10} \begin{array}{*{20}{l}}
		{R_{mn} =  - \frac{1}{2}\sum\limits_{p =  - \infty }^{ + \infty } {} [2i\alpha '{\pi ^2}\left( {\gamma _{p - n}^{\nu \mu } + \gamma _{m - p}^{\mu \nu }} \right)}
		{ + \frac{i}{{2\alpha '}}\left( {{{\left( {p - n} \right)}^2}\theta _{p - n}^{\nu \mu } + {{\left( {m - p} \right)}^2}\theta _{m - p}^{\mu \nu }} \right)\left. {} \right]\alpha _p^\mu \alpha _{m + n - p}^\nu }
	\end{array}
\end{equation}
which is not the same result given by S-Z Mousavi in \cite{bib10}.\\
The super-algebra then, is given by:\\
For N-S sector:
\begin{equation}
	\label{15}	[L_m,L_n] = (m-n)L_{n+m} + \frac{D}{{8}}m\left( {{m^2} - 1} \right){\delta _{m + n,0}} + R_{mn}
\end{equation}
\begin{equation}
	\label{lg}	\left[ {{L_m},{G_r}} \right] = \left( {\frac{1}{2}m - r} \right){G_{m + r}}+ V_{mr}
\end{equation}
\begin{equation}
	\label{gg}	\left\{ {{G_r},{G_s}} \right\} = 2{L_{r + s}} + \frac{D}{2}\left( {{r^2} - \frac{1}{4}} \right){\delta _{r + s}} + B_{rs}
\end{equation}
with the new anomaly terms  $B_{rs}$, $V_{mr}$ given by : 
\begin{equation}
	\label{bb} B_{rs} =  - \frac{1}{2}\sum\limits_{q =  - \infty }^{ + \infty } {\left[ \begin{array}{l}
			2i\alpha '{\pi ^2}\left( {\gamma _{q - s}^{\nu \mu } + \gamma _{r - q}^{\mu \nu }} \right) + 
			\frac{i}{{2\alpha '}}\left( \begin{array}{l}
				{\left( {q - s} \right)^2}\theta _{q - s}^{\nu \mu }
				+ {\left( {r - q} \right)^2}\theta _{r - q}^{\mu \nu }
			\end{array} \right)
		\end{array} \right]b_q^\mu b_{r + s - q}^\nu }  
\end{equation}
\begin{equation}
	\label{vv} V_{mr} =  - \frac{1}{2}\sum\limits_{q =  - \infty }^{ + \infty } {\left[ \begin{array}{l}
			2i\alpha '{\pi ^2}\left( {\gamma _{q - r}^{\nu \mu } + \gamma _{m - q}^{\mu \nu }} \right) + 
			\frac{i}{{2\alpha '}}\left( \begin{array}{l}
				{\left( {q - r} \right)^2}\theta _{q - r}^{\nu \mu }
				+ {\left( {m - q} \right)^2}\theta _{m - q}^{\mu \nu }
			\end{array} \right)
		\end{array} \right]\alpha _q^\mu b_{r + m - q}^\nu }
\end{equation}
{where $n \in \mathbb{Z}$ and $r \in \mathbb{Z} + \frac{1}{2}$. In $B_{rs}$, $q$ sums over all half-integer modes while in $V_{mr}$, $q$ sums over all integer modes.\\}
For Ramond sector:
\begin{equation}
	\label{18}	[L_m,L_n] = (m-n)L_{n+m} + \frac{D}{{8}}{m^{3}} {\delta _{m + n,0}} + R_{mn}
\end{equation}
\begin{equation}
	\label{19}	\left[ {{L_m},{F_n}} \right] = \left( {\frac{1}{2}m - n} \right){F_{m + n}}+ W_{mn}
\end{equation}
\begin{equation}
	\label{20}	\left\{ {{F_r},{F_s}} \right\} = 2{L_{r + s}} + \frac{D}{2}r^2{\delta _{r + s}} + D_{rs}
\end{equation}
with again, the new anomaly terms  $D_{rs}$, $W_{mn}$ given by : 
\begin{equation}
	\label{dd} D_{rs} =  - \frac{1}{2}\sum\limits_{q =  - \infty }^{ + \infty } {\left[ \begin{array}{l}
			2i\alpha '{\pi ^2}\left( {\gamma _{q - s}^{\nu \mu } + \gamma _{r - q}^{\mu \nu }} \right) + 
			\frac{i}{{2\alpha '}}\left( \begin{array}{l}
				{\left( {q - s} \right)^2}\theta _{q - s}^{\nu \mu }
				+ {\left( {r - q} \right)^2}\theta _{r - q}^{\mu \nu }
			\end{array} \right)
		\end{array} \right]d_q^\mu d_{r + s - q}^\nu } 
\end{equation}
\begin{equation}
	\label{ww} W_{mn} =  - \frac{1}{2}\sum\limits_{q =  - \infty }^{ + \infty } {\left[ \begin{array}{l}
			2i\alpha '{\pi ^2}\left( {\gamma _{q - n}^{\nu \mu } + \gamma _{m - q}^{\mu \nu }} \right) + 
			\frac{i}{{2\alpha '}}\left( \begin{array}{l}
				{\left( {q - n} \right)^2}\theta _{q - n}^{\nu \mu }
				+ {\left( {m - q} \right)^2}\theta _{m - q}^{\mu \nu }
			\end{array} \right)
		\end{array} \right]\alpha _q^\mu d_{n + m - q}^\nu } 
\end{equation}
{where $n,r \in \mathbb{Z}$ and $q$ sums over all integer modes.}
\section{Modified Lorentz Algebra}\label{sec4}
The angular momentum $M^{\mu\nu}$ is given by \cite{bib15,bib14,bib16,bib9}:
\begin{equation}
	\label{98} {{M^{\mu \nu }} = \left\{ {\begin{array}{*{20}{l}}
				\begin{array}{l}
					{x^\mu }{p^\nu } - {x^\nu }{p^\mu } - i\sum\limits_{n = 1}^\infty  {\frac{1}{n}\left( {\alpha _{ - n}^\mu \alpha _n^\nu  - \alpha _{ - n}^\nu \alpha _n^\mu } \right)}  - 
					\frac{i}{4}\sum\limits_{r =  - \infty }^{ + \infty } {\left( {b_{ - r}^\mu b_r^\nu  - b_{ - r}^\nu b_r^\mu } \right)}  \to {\rm{N - S}}{\rm{sector}}
				\end{array}\\
				\begin{array}{l}
					{x^\mu }{p^\nu } - {x^\nu }{p^\mu } - i\sum\limits_{n = 1}^\infty  {\frac{1}{n}\left( {\alpha _{ - n}^\mu \alpha _n^\nu  - \alpha _{ - n}^\nu \alpha _n^\mu } \right)}  - 
					\frac{i}{4}\sum\limits_{m =  - \infty }^{ + \infty } {\left( {d_{ - m}^\mu d_m^\nu  - d_{ - m}^\nu d_m^\mu } \right)}  \to {\rm{R}}{\rm{sector}}
				\end{array}
		\end{array}} \right.}
\end{equation}
By the use of the equations (\ref{66}) (\ref{6}) (\ref{669}), a direct calculation (see Appendix \ref{222}) gives the following modified Lorentz algebra:
\begin{equation}
	\label{nn} \begin{array}{l}
		[{M^{\mu \nu }},{M^{\rho \lambda }}] =  - i{\eta ^{\nu \rho }}{M^{\mu \lambda }} + i{\eta ^{\mu \lambda }}{M^{\rho \nu }} +
		i{\eta ^{\nu \lambda }}{M^{\mu \rho }} - i{\eta ^{\mu \rho }}{M^{\lambda \nu }} + {T^{\mu \nu \rho \lambda }}
	\end{array}
\end{equation}
\begin{equation}
	\label{xx} 
	[{p^\mu },{M^{\nu \rho }}] = i{\eta ^{\rho \mu }}{p^\nu } - i{\eta ^{\nu \mu }}{p^\rho } + K^{\nu\mu\rho}
\end{equation}
\begin{equation}
	\label{yy}{\left[ {{p^\mu },{p^\nu }} \right] = i{\pi ^2}{\gamma _0}^{\mu \nu }}
\end{equation}
Where $T^{\mu\nu\rho\lambda}$, $K^{\nu\mu\rho}$ represent the anomalies due to  the non-commutativity and which are given by:
\begin{equation}
	\label{99}\begin{array}{*{20}{l}}
		{{T^{\mu \nu \rho \lambda }} = i{\pi ^2}\left( {\begin{array}{*{20}{l}}
					{\gamma _0^{\nu \lambda }{x^\mu }{x^\rho } + \gamma _0^{\nu \rho }{x^\mu }{x^\lambda } + }\\
					{\gamma _0^{\mu \lambda }{x^\nu }{x^\rho } + \gamma _0^{\mu \rho }{x^\nu }{x^\lambda }}
			\end{array}} \right) + 2i{\pi ^2}\alpha '\tau \left( {\begin{array}{*{20}{l}}
					{\gamma _0^{\nu \rho }{x^\mu }{p^\lambda } - \gamma _0^{\mu \lambda }{x^\rho }{p^\nu } + }\\
					{\gamma _0^{\nu \lambda }{x^\mu }{p^\rho } - \gamma _0^{\mu \rho }{x^\lambda }{p^\nu } + }\\
					{\gamma _0^{\mu \rho }{x^\nu }{p^\lambda } - \gamma _0^{\nu \lambda }{x^\rho }{p^\mu } + }\\
					{\gamma _0^{\mu \lambda }{x^\nu }{p^\rho } - \gamma _0^{\nu \rho }{x^\lambda }{p^\mu }}
			\end{array}} \right) + }\\
		{\left( {i\theta _0^{\mu \rho } - 4i{\pi ^2}{{\alpha '}^2}{\tau ^2}\gamma _0^{\mu \rho }} \right){p^\lambda }{p^\nu } + \left( {i\theta _0^{\mu \lambda } - 4i{\pi ^2}{{\alpha '}^2}{\tau ^2}\gamma _0^{\mu \lambda }} \right){p^\rho }{p^\nu } + }\\
		{\left( {i\theta _0^{\nu \rho } - 4i{\pi ^2}{{\alpha '}^2}{\tau ^2}\gamma _0^{\nu \rho }} \right){p^\lambda }{p^\mu } + \left( {i\theta _0^{\nu \lambda } - 4i{\pi ^2}{{\alpha '}^2}{\tau ^2}\gamma _0^{\nu \lambda }} \right){p^\rho }{p^\mu } + }\\
		{\left( {i\frac{{{{(2\pi \alpha ')}^2}}}{{2\alpha '}}{\gamma _n}^{\nu \rho } + i\frac{{{n^2}}}{{2\alpha '}}{\theta _n}^{\nu \rho }} \right)\left( {\alpha _{ - n}^\mu \alpha _n^\lambda  + \alpha _{ - n}^\lambda \alpha _n^\mu } \right) + \left( {i\frac{{{{(2\pi \alpha ')}^2}}}{{2\alpha '}}{\gamma _n}^{\mu \lambda } + i\frac{{{n^2}}}{{2\alpha '}}{\theta _n}^{\mu \lambda }} \right)\left( {\alpha _{ - n}^\rho \alpha _n^\lambda  + \alpha _{ - n}^\lambda \alpha _n^\rho } \right) + }\\
		{\left( {i\frac{{{{(2\pi \alpha ')}^2}}}{{2\alpha '}}{\gamma _n}^{\nu \lambda } + i\frac{{{n^2}}}{{2\alpha '}}{\theta _n}^{\nu \lambda }} \right)\left( {\alpha _{ - n}^\rho \alpha _n^\mu  + \alpha _{ - n}^\mu \alpha _n^\rho } \right) + \left( {i\frac{{{{(2\pi \alpha ')}^2}}}{{2\alpha '}}{\gamma _n}^{\mu \rho } + i\frac{{{n^2}}}{{2\alpha '}}{\theta _n}^{\mu \rho }} \right)\left( {\alpha _{ - n}^\lambda \alpha _n^\nu  + \alpha _{ - n}^\nu \alpha _n^\lambda } \right)}\\
	\end{array}
\end{equation}
\begin{equation}
	\label{ee}
	\begin{array}{l}
		{K^{\nu \mu \rho }} = 2i{\pi ^2}\alpha '\tau \gamma _0^{\nu \mu }{p^\rho } - 2i{\pi ^2}\alpha '\tau \gamma _0^{\rho \mu }{p^\nu }
		+ i{\pi ^2}\gamma _0^{\mu \rho }{p^\nu } - i{\pi ^2}\gamma _0^{\mu \nu }{p^\rho }
	\end{array}
\end{equation}

\section{Mass Spectrum and GSO Projection} 
The calculation of the mass spectrum required working in the light cone coordinates. The equation (\ref{6}) will take this form:
\begin{equation}
	\label{22} \left[ {\alpha _m^i ,\alpha _n^j } \right] = \left( {m{\eta ^{ij }} + i \frac{{{(2\pi \alpha ')^2}}}{{2\alpha '}}{\gamma _n}^{ij } + i \frac{{{n^2}}}{{2\alpha '}}{\theta _n}^{ij }} \right){\delta _{n + m,0}}
\end{equation}
where $i,j =2..D-1$ represents the transversal indices, while the anti-commutation relations between the fermionic modes remain unchanged.
\begin{equation}
	\left\{ {d_m^i,d_n^j} \right\} =\eta^{ij} \delta_{m+n,0}
\end{equation}
\begin{equation}
	\left\{ {b_r^i,b_s^j} \right\} =\eta^{ij} \delta_{r+s,0}
\end{equation}
The mass operator for both sectors are given by:\\
.Ramond sector:
\begin{equation}
	\label{23}{M^2_R} = \frac{1}{{\alpha '}}\left( {\sum\limits_{n = 1}^\infty  {\alpha _{ - n}^i\alpha _n^i}  + \sum\limits_{r = 1}^\infty  {rd_{ - r}^id_r^i} } \right)
\end{equation}
.Neveu-Schwarz sector:
\begin{equation}
	\label{24}{M^2_{NS}} = \frac{1}{{\alpha '}}\left( {\sum\limits_{n = 1}^\infty  {\alpha _{ - n}^i\alpha _n^i}  + \sum\limits_{r = \frac{1}{2}}^\infty  {rb_{ - r}^ib_r^i - \frac{1}{2}} } \right)
\end{equation}
We need to diagonalize the antisymmetric matrices $\theta _m^{ij}$ and $\gamma _m^{ij}$ by introducing the unitary matrix $U_m$ such that:
\begin{equation}
	\label{25} {\left( {{U^{ - 1}_{m}}i{\theta _m}U_{m}} \right)^{ij}} = D_m^{ij} = \mu _i^{\left( m \right)}{\delta ^{ij}}
\end{equation}
and,
\begin{equation}
	\label{26} {\left( {{U^{ - 1}_{m}}i{\gamma _m}U_{m}} \right)^{ij}} = T_m^{ij} = \nu _i^{\left( m \right)}{\delta ^{ij}}
\end{equation}
With $\left[ {{\theta _m},{\gamma _m}} \right] = 0$ and $\mu _i^{\left( m \right)}$ and $\nu _i^{\left( m \right)}$ are the eigenvalues of $i{\theta _m} ^{ij}$ and $i{\gamma _m} ^{ij}$ respectively\\
This latter diagonalization, given by equations \eqref{25} and \eqref{26}, can be obtained through a redefinition of the Fock space \cite{bib13} in order to get a diagonal mass in this new basis. The redefinition take this form:
\begin{equation}
	\label{21} \begin{array}{*{20}{l}}
		{\prod\limits_{i = 2}^{D - 1} {\prod\limits_{m = 1}^\infty  {{{\left( {\alpha _{ - m}^i} \right)}^{{\lambda _{m,i}}}}\prod\limits_{j = 2}^{D - 1} {\left( \begin{array}{l}
							\prod\limits_{r = \frac{1}{2},\frac{3}{2}...} {{{\left( {b_{ - r}^j} \right)}^{{\rho _{r,j}}}}} \\
							or\\
							\prod\limits_{n = 1,2..} {{{\left( {d_{ - n}^j} \right)}^{{\rho _{n,j}}}}} 
						\end{array} \right)} } } \vert {p^ + },{{\vec p}^T}\rangle  \to }\\
		{\prod\limits_{i = 2}^{D - 1} {\prod\limits_{m = 1}^\infty  {{{\left\{ {{{\left( {U_m^{ - 1}{\alpha _{ - m}}} \right)}^i}} \right\}}^{{\lambda _{m,i}}}}} } \prod\limits_{j = 2}^{D - 1} {\left( \begin{array}{l}
					\prod\limits_{r = \frac{1}{2},\frac{3}{2}...} {{{\left( {b_{ - r}^j} \right)}^{{\rho _{r,j}}}}} \\
					or\\
					\prod\limits_{n = 1,2..} {{{\left( {d_{ - n}^j} \right)}^{{\rho _{n,j}}}}} 
				\end{array} \right)} \vert {p^ + },{{\vec p}^T}\rangle }
	\end{array}
\end{equation}
Where the non negative integer $\lambda _{m,i}$ shows \cite{bib15} how many times the creation operator $\alpha _{ - m}^i$ appears, and $\rho _{n,k}$ takes either zero or one.\\ 
In order to get an equivalent of a GSO projection, one can use the usual way to get the following steps in (Table \ref{tab1}) and (Table \ref{tab2}).\\
The results of GSO projection for the two sectors are grouped in (Table \ref{tab3}).\\
\clearpage
\onecolumn
\begin{table}[h]
	\medskip
	\centering\renewcommand{\arraystretch}{1.2}
	\begin{tabular}{{l c c}}
		\toprule
		Level & \multicolumn{2}{c}{N-S Sector} \\
		\midrule
		&state & Mass \\
		\midrule
		0 &	$\vert 0 \rangle $ & $- \frac{1}{{2\alpha '}}$\\
		\midrule
		{1} & $b^{i}_{-\frac{1}{2}}\vert 0 \rangle $ & 0\\
		\midrule
		{2} & $b_{ - \frac{1}{2}}^ib_{ - \frac{1}{2}}^j\vert 0 \rangle$ & $\frac{1}{{2\alpha '}}$\\
		& ${U^{ - 1}_{1}}\alpha _{ - 1}^i\vert 0 \rangle$ &$  \frac{1}{\alpha '}\left( \frac{1}{2} -\frac{1}{2{\alpha '}} \left( {\mu _j^{\left( 1 \right)} +  {{\left( {2\pi \alpha '} \right)}^2} \nu _j^{\left( 1 \right)}} \right)\right)$ \\
		\midrule
		{3} &$b_{ - \frac{1}{2}}^ib_{ - \frac{1}{2}}^jb_{ - \frac{1}{2}}^k\vert 0 \rangle$ & $\frac{1}{{\alpha '}}$\\
		& $b_{ - \frac{3}{2}}^i\vert 0 \rangle$ &$\frac{1}{{\alpha '}}$\\
		& ${U^{ - 1}_{1}}\alpha _{ - 1}^ib_{ - \frac{1}{2}}^j\vert 0 \rangle$ &$  \frac{1}{\alpha '}\left( 1 -\frac{1}{2{\alpha '}} \left( {\mu _j^{\left( 1 \right)} +  {{\left( {2\pi \alpha '} \right)}^2} \nu _j^{\left( 1 \right)}} \right)\right)$ \\
		\midrule
		{4}
		& $b_{ - \frac{1}{2}}^ib_{ - \frac{1}{2}}^jb_{ - \frac{1}{2}}^k b_{ - \frac{1}{2}}^l \vert 0 \rangle$ & $ \frac{3}{{2\alpha '}}$\\
		& $b_{ - \frac{3}{2}}^ib_{ - \frac{1}{2}}^j \vert 0 \rangle$ & $ \frac{3}{{2\alpha '}}$\\
		& ${U^{ - 1}_{2}}\alpha _{ - 2}^i\vert 0 \rangle$&  $  \frac{1}{\alpha '}\left( \frac{3}{2} - \frac{1}{2{\alpha '}}\left( 4 {\mu _j^{\left( 2 \right)} +  {\left( {2\pi \alpha '} \right)}^2} \nu _j^{\left( 2 \right)} \right)\right)$\\
		& ${U^{ - 1}_{1}}\alpha _{ - 1}^j{U^{ - 1}_{1}}\alpha _{ - 1}^k\vert 0 \rangle$ & $  \frac{1}{\alpha '}\left( \frac{3}{2} - \frac{1}{2{\alpha '}}\left(  {\mu _j^{\left( 1 \right)} +  {\left( {2\pi \alpha '} \right)}^2} \nu _j^{\left( 1 \right)} + {\mu _k^{\left( 1 \right)} +  {\left( {2\pi \alpha '} \right)}^2} \nu _k^{\left( 1 \right)} \right)\right)$\\
		& ${U^{ - 1}_{1}}\alpha _{ - 1}^jb_{ - \frac{1}{2}}^kb_{ - \frac{1}{2}}^l\vert 0 \rangle$ &  $  \frac{1}{\alpha '}\left( \frac{3}{2} - \frac{1}{2{\alpha '}}\left(  {\mu _j^{\left( 1 \right)} +  {\left( {2\pi \alpha '} \right)}^2} \nu _j^{\left( 1 \right)} \right)\right)$\\
		\midrule
		{5} 
		& $b_{ - \frac{1}{2}}^ib_{ - \frac{1}{2}}^jb_{ - \frac{1}{2}}^k b_{ - \frac{1}{2}}^lb_{ - \frac{1}{2}}^m \vert 0 \rangle$ & $\frac{2}{{\alpha '}}$\\
		& $b_{ - \frac{5}{2}}^i\vert 0 \rangle$ & $\frac{2}{{\alpha '}}$\\
		& $b_{ - \frac{1}{2}}^i b_{ - \frac{1}{2}}^jb_{ - \frac{1}{2}}^k$ & $\frac{2}{{\alpha '}}$\\
		& ${U^{ - 1}_{2}}\alpha _{ - 2}^jb_{ - \frac{1}{2}}^k\vert 0 \rangle$&  $  \frac{1}{\alpha '}\left(2 - \frac{1}{2{\alpha '}}\left( 4 {\mu _j^{\left( 2 \right)} +  {\left( {2\pi \alpha '} \right)}^2} \nu _j^{\left( 2 \right)} \right)\right)$\\
		& ${U^{ - 1}_{1}}\alpha _{ - 1}^j{U^{ - 1}_{1}}\alpha _{ - 1}^kb_{ - \frac{1}{2}}^l\vert 0 \rangle$ & $  \frac{1}{\alpha '}\left(2- \frac{1}{2{\alpha '}}\left(  {\mu _j^{\left( 1 \right)} +  {\left( {2\pi \alpha '} \right)}^2} \nu _j^{\left( 1 \right)} + {\mu _k^{\left( 1 \right)} +  {\left( {2\pi \alpha '} \right)}^2} \nu _k^{\left( 1 \right)} \right)\right)$\\
		& ${U^{ - 1}_{ 1}}\alpha _{ - 1}^jb_{ - \frac{1}{2}}^kb_{ - \frac{1}{2}}^lb_{ - \frac{1}{2}}^m\vert 0 \rangle$ &  $  \frac{1}{\alpha '}\left(2 - \frac{1}{2{\alpha '}}\left(  {\mu _j^{\left( 1 \right)} +  {\left( {2\pi \alpha '} \right)}^2} \nu _j^{\left( 1 \right)} \right)\right)$\\
		& ${U^{ - 1}_{1}}\alpha _{ - 1}^jb_{ - \frac{3}{2}}^k\vert 0 \rangle$ &  $  \frac{1}{\alpha '}\left(2 - \frac{1}{2{\alpha '}}\left(  {\mu _j^{\left( 1 \right)} +  {\left( {2\pi \alpha '} \right)}^2} \nu _j^{\left( 1 \right)} \right)\right)$\\
		\bottomrule
	\end{tabular}
	\caption{This table represent the mass spectrum in terms of redefined modes.}
	\label{tab1}
\end{table}
\begin{table}[h] 
	\medskip
	\centering\renewcommand{\arraystretch}{1.2}
	\begin{tabular}{l c c}
		\toprule
		Level & \multicolumn{2}{c}{R-Sector} \\
		\midrule
		&state & Mass\\
		\midrule
		0 & $\vert 0 \rangle $ & 0\\
		\midrule
		{1} &$d^{j}_{-1}\vert 0 \rangle $ & $\frac{1}{\alpha\prime}$\\
		& ${U^{ - 1}_{ 1}}\alpha _{ - 1}^j\vert 0 \rangle$ & $  \frac{1}{\alpha '}\left( 1 - \left(\frac{1}{2{\alpha '}} {\mu _j^{\left( 1 \right)} +  \frac{{{\left( {2\pi \alpha '} \right)}^2}} {2{\alpha '}}\nu _j^{\left( 1 \right)}} \right)\right)$\\
		\midrule
		{2} 
		& $d_{ - 2}^j\vert 0 \rangle$ & $\frac{2}{{\alpha '}}$\\
		& $d_{ - 1}^jd_{ - 1}^k\vert 0 \rangle$ &  $\frac{2}{{\alpha '}}$\\
		& ${U^{ - 1}_{ 2}}\alpha _{ - 2}^j\vert 0 \rangle$ &$  \frac{1}{\alpha '}\left(2 - \frac{1}{2{\alpha '}}\left( 4 {\mu _j^{\left( 2 \right)} +  {\left( {2\pi \alpha '} \right)}^2} \nu _j^{\left( 2 \right)} \right)\right)$\\
		& $\left( {{U^{ - 1}_{ 1}}\alpha _{ - 1}^j} \right)\left( {{U^{ - 1}_{ 1}}\alpha _{ - 1}^k} \right)\vert 0 \rangle $ & $\frac{1}{{\alpha '}}\left( {2 - \frac{1}{{2\alpha '}}\left( \begin{array}{l}
				\mu _j^{\left( 1 \right)} + {\left( {2\pi \alpha '} \right)^2}\nu _j^{\left( 1 \right)} + \\
				\mu _k^{\left( 1 \right)} + {\left( {2\pi \alpha '} \right)^2}\nu _k^{\left( 1 \right)}
			\end{array} \right)} \right)$\\
		& $\left( {{U^{ - 1}_{ 1}}\alpha _{ - 1}^j} \right)d_{ - 1}^k\vert 0 \rangle$ & $  \frac{1}{\alpha '}\left(2- \frac{1}{2{\alpha '}}\left(  {\mu _j^{\left( 1 \right)} +  {\left( {2\pi \alpha '} \right)}^2} \nu _j^{\left( 1 \right)} \right)\right)$\\
		\bottomrule
	\end{tabular}
	\caption{Note: This table represent the mass spectrum in terms of redefined modes.}
	\label{tab2}
\end{table}
\newpage
Then, one can impose :\\
\begin{equation}
	\label{40} \nu _i^{\left( 1 \right)} =   \frac{-1}{\left(2\pi{\alpha '}\right)^2}\mu _i^{\left( 1 \right)}
\end{equation}
to restore the value of the mass for the first excited state (for example), and in general:
\begin{equation}
	\label{41} \nu _i^{\left( m \right)} =   \frac{-m^2}{\left(2\pi{\alpha '}\right)^2}\mu _i^{\left( m \right)}
\end{equation}
being an equivalent to :
\begin{equation}
	\label{43} T^{ij}_{(m)} =\frac{-m^2}{\left(2\pi{\alpha '}\right)^2}D^{ij}_{(m)}  
\end{equation}
which can be use to restore mass values of the other levels, where $m>0$ represent the number of state level.\\
In the present construction, the non-commutative deformation is implemented through a redefinition of the bosonic oscillators, while the fermionic modes remain unchanged. Consequently, the fermionic Fock space and its canonical anticommutation relations are preserved. The supersymmetric pairing underlying the GSO projection therefore remains intact within the considered perturbative regime.\\
\clearpage
\begin{landscape}
	\begin{table}[p]   % [p] = page dédiée
		\centering
		\footnotesize   % ou \scriptsize si besoin encore plus petit
		\renewcommand{\arraystretch}{0.88}   % resserre verticalement
		
		\resizebox{\paperwidth}{!}{%
	\begin{tabular}{l c c c c}
		\toprule
		Level & \multicolumn{2}{c}{N-S Sector} & \multicolumn{2}{c}{R-Sector} \\
		\midrule
		&state & Mass&state & Mass\\
		\midrule
		{1} & $b^{i}_{-\frac{1}{2}}\vert 0 \rangle$ & $0$ & $\vert 0 \rangle$ & $0$\\
		\midrule
		{3} & $b_{ - \frac{1}{2}}^i b_{ - \frac{1}{2}}^j b_{ - \frac{1}{2}}^k\vert 0 \rangle$ & $\frac{1}{{\alpha '}}$ & $d_{ - 1}^i\vert 0 \rangle$ & $\frac{1}{{\alpha '}}$ \\
		& $b_{ - \frac{3}{2}}^i\vert 0 \rangle$ & $\frac{1}{{\alpha '}}$ & & \\
		& ${U^{ - 1}_{ 1}}\alpha _{ - 1}^ib_{ - \frac{1}{2}}^j\vert 0 \rangle$ & $  \frac{1}{\alpha '}\left( 1 + \left(\frac{1}{2{\alpha '}} {\mu _j^{\left( 1 \right)} +  \frac{{{\left( {2\pi \alpha '} \right)}^2}} {2{\alpha '}}\nu _j^{\left( 1 \right)}} \right)\right)$ & ${U^{ - 1}_{ 1}}\alpha _{ - 1}^i\vert 0 \rangle$ & $  \frac{1}{\alpha '}\left( 1 + \left(\frac{1}{2{\alpha '}} {\mu _j^{\left( 1 \right)} +  \frac{{{\left( {2\pi \alpha '} \right)}^2}} {2{\alpha '}}\nu _j^{\left( 1 \right)}} \right)\right)$ \\
		\midrule
		{5} & $b_{ - \frac{1}{2}}^ib_{ - \frac{1}{2}}^jb_{ - \frac{1}{2}}^k b_{ - \frac{1}{2}}^lb_{ - \frac{1}{2}}^m \vert 0 \rangle$  & $\frac{2}{{\alpha '}}$	& $d_{ - 2}^j\vert 0 \rangle$ & $\frac{2}{{\alpha '}}$\\
		& $b_{ - \frac{5}{2}}^i\vert 0 \rangle$ & $\frac{2}{{\alpha '}}$ & $d_{ - 1}^jd_{ - 1}^k\vert 0 \rangle$ &  $\frac{2}{{\alpha '}}$\\
		& $b_{ - \frac{1}{2}}^i b_{ - \frac{1}{2}}^jb_{ - \frac{1}{2}}^k$ & $\frac{2}{{\alpha '}}$\\
		& ${U^{ - 1}_{ 2}}\alpha _{ - 2}^jb_{ - \frac{1}{2}}^k\vert 0 \rangle$&  $  \frac{1}{\alpha '}\left(2 - \frac{1}{2{\alpha '}}\left( 4 {\mu _j^{\left( 2 \right)} +  {\left( {2\pi \alpha '} \right)}^2} \nu _j^{\left( 2 \right)} \right)\right)$& ${U^{ - 1}_{ 2}}\alpha _{ - 2}^j\vert 0 \rangle$ &$  \frac{1}{\alpha '}\left(2 - \frac{1}{2{\alpha '}}\left( 4 {\mu _j^{\left( 2 \right)} +  {\left( {2\pi \alpha '} \right)}^2} \nu _j^{\left( 2 \right)} \right)\right)$\\
		& ${U^{ - 1}_{ 1}}\alpha _{ - 1}^j{U^{ - 1}_{ 1}}\alpha _{ - 1}^kb_{ - \frac{1}{2}}^l\vert 0 \rangle$ & $\frac{1}{{\alpha '}}\left( {2 - \frac{1}{{2\alpha '}}\left( \begin{array}{l}
				\mu _j^{\left( 1 \right)} + {\left( {2\pi \alpha '} \right)^2}\nu _j^{\left( 1 \right)} + \\
				\mu _k^{\left( 1 \right)} + {\left( {2\pi \alpha '} \right)^2}\nu _k^{\left( 1 \right)}
			\end{array} \right)} \right)$& $\left( {{U^{ - 1}_{ 1}}\alpha _{ - 1}^j} \right)\left( {{U^{ - 1}_{ 1}}\alpha _{ - 1}^k} \right)\vert 0 \rangle $ &$\frac{1}{{\alpha '}}\left( {2 - \frac{1}{{2\alpha '}}\left( \begin{array}{l}
				\mu _j^{\left( 1 \right)} + {\left( {2\pi \alpha '} \right)^2}\nu _j^{\left( 1 \right)} + \\
				\mu _k^{\left( 1 \right)} + {\left( {2\pi \alpha '} \right)^2}\nu _k^{\left( 1 \right)}
			\end{array} \right)} \right)$\\
		& ${U^{ - 1}_{ 1}}\alpha _{ - 1}^jb_{ - \frac{1}{2}}^kb_{ - \frac{1}{2}}^lb_{ - \frac{1}{2}}^m\vert 0 \rangle$ &  $  \frac{1}{\alpha '}\left(2 - \frac{1}{2{\alpha '}}\left(  {\mu _j^{\left( 1 \right)} +  {\left( {2\pi \alpha '} \right)}^2} \nu _j^{\left( 1 \right)} \right)\right)$\\
		& ${U^{ - 1}_{ 1}}\alpha _{ - 1}^jb_{ - \frac{3}{2}}^k\vert 0 \rangle$ &  $  \frac{1}{\alpha '}\left(2 - \frac{1}{2{\alpha '}}\left(  {\mu _j^{\left( 1 \right)} +  {\left( {2\pi \alpha '} \right)}^2} \nu _j^{\left( 1 \right)} \right)\right)$& $\left( {{U^{ - 1}_{ 1}}\alpha _{ - 1}^j} \right)d_{ - 1}^k\vert0 \rangle$ & $  \frac{1}{\alpha '}\left(2- \frac{1}{2{\alpha '}}\left(  {\mu _j^{\left( 1 \right)} +  {\left( {2\pi \alpha '} \right)}^2} \nu _j^{\left( 1 \right)} \right)\right)$\\
		\bottomrule
	\end{tabular}}
	\caption{This table represent the GSO projection for the two sectors.}
	\label{tab3}
\end{table}
\end{landscape}
\clearpage
Under the imposed condition on the deformation parameters \ref{41}, the resulting mass spectrum coincides with the commutative one at first order of $\theta$ (Table \ref{tab4}). Beyond this order, higher-order corrections are expected to modify the spectrum, and a full analysis of these effects lies outside the scope of the present work.\\

\begin{table}[h]
	\medskip
	\centering\renewcommand{\arraystretch}{1.2}
	\begin{tabular}{l c c c c}
		\toprule
		Level & \multicolumn{2}{c}{N-S Sector} & \multicolumn{2}{c}{R-Sector} \\
		\midrule
		&state & Mass&state & Mass\\
		\midrule
		{1} & $b^{i}_{-\frac{1}{2}}\vert 0 \rangle$ & $0$ & $\vert 0 \rangle$ & $0$\\
		\midrule
		{3} & $b_{ - \frac{1}{2}}^ib_{ - \frac{1}{2}}^jb_{ - \frac{1}{2}}^k\vert 0 \rangle$ & $\frac{1}{{\alpha '}}$ & $d_{ - 1}^i\vert 0 \rangle$ & $\frac{1}{{\alpha '}}$ \\
		& $b_{ - \frac{3}{2}}^i\vert 0 \rangle$ & $\frac{1}{{\alpha '}}$ & & \\
		& ${U^{ - 1}_{ 1}}\alpha _{ - 1}^ib_{ - \frac{1}{2}}^j\vert 0 \rangle$ & $\frac{1}{{\alpha '}}$ & ${U^{ - 1}_{ 1}}\alpha _{ - 1}^i\vert 0 \rangle$ & $  \frac{1}{{\alpha '}}$ \\
		\hline
		{5} & $b_{ - \frac{1}{2}}^ib_{ - \frac{1}{2}}^jb_{ - \frac{1}{2}}^k b_{ - \frac{1}{2}}^lb_{ - \frac{1}{2}}^m \vert0 \rangle$  & $\frac{2}{{\alpha '}}$	& $d_{ - 2}^j\vert 0 \rangle$ & $\frac{2}{{\alpha '}}$\\
		& $b_{ - \frac{5}{2}}^i\vert 0 \rangle$ & $\frac{2}{{\alpha '}}$ & $d_{ - 1}^jd_{ - 1}^k\vert 0 \rangle$ &  $\frac{2}{{\alpha '}}$\\
		& $b_{ - \frac{1}{2}}^i b_{ - \frac{1}{2}}^jb_{ - \frac{1}{2}}^k$ & $\frac{2}{{\alpha '}}$\\
		& ${U^{ - 1}_{ 2}}\alpha _{ - 2}^jb_{ - \frac{1}{2}}^k\vert 0 \rangle$&  $  \frac{2}{{\alpha '}}$ & ${U^{ - 1}_{ 2}}\alpha _{ - 2}^j\vert 0 \rangle$ &$\frac{2}{{\alpha '}}$\\
		& ${U^{ - 1}_{ 1}}\alpha _{ - 1}^j{U^{ - 1}_{ 1}}\alpha _{ - 1}^kb_{ - \frac{1}{2}}^l\vert 0 \rangle$ & $\frac{2}{{\alpha '}}$ & $\left( {{U^{ - 1}_{ 1}}\alpha _{ - 1}^j} \right)\left( {{U^{ - 1}_{ 1}}\alpha _{ - 1}^k} \right)\vert 0 \rangle $ &$\frac{2}{{\alpha '}}$\\
		& ${U^{ - 1}_{ 1}}\alpha _{ - 1}^jb_{ - \frac{1}{2}}^kb_{ - \frac{1}{2}}^lb_{ - \frac{1}{2}}^m\vert 0 \rangle$ &  $  \frac{2}{{\alpha '}}$\\
		& ${U^{ - 1}_{ 1}}\alpha _{ - 1}^jb_{ - \frac{3}{2}}^k\vert 0 \rangle$ &  $\frac{2}{{\alpha '}}$& $\left( {{U^{ - 1}_{ 1}}\alpha _{ - 1}^j} \right)d_{ - 1}^k\vert 0 \rangle$ & $\frac{2}{{\alpha '}}$\\
		\hline
	\end{tabular}
	\caption{This table represent the first levels of the mass spectrum after GSO projection and the application of the equation (\ref{41}).}
	\label{tab4}
\end{table}

By applying ($U_{m} U^{-1}_{m}$) on the both sides of equations (\ref{25}) and (\ref{26}), one can show that the equation (\ref{43}) can be expressed with respect to $\theta^{ij}_{m}$ and $\gamma^{ij}_m$ as
\begin{equation}
	\label{456} \gamma^{ij}_{(m)} =\frac{-m^2}{\left(2\pi{\alpha '}\right)^2} \theta^{ij}_{(m)}  
\end{equation}
From this result, we can fix our starting model (\ref{1}) by imposing to $\theta^{\mu\nu}$ and $\gamma^{\mu\nu}$ the following relation:
\begin{equation}
	\label{455} \gamma^{\mu\nu}_{(m)} =\frac{-m^2}{\left(2\pi{\alpha '}\right)^2} \theta^{\mu\nu}_{(m)}  
\end{equation}
where $m\ne 0$ and $\mu,\nu = 0,1...,D-1$. \\
With this condition (\ref{455}), one can easily verify that all the anomaly terms (\ref{10}), (\ref{bb}), (\ref{vv}), (\ref{dd}) and (\ref{ww}) of the modified Virasoro algebra due to the non-commutativity are eliminated. This result is a direct consequence of the fact that we considered non-commutativity between coordiantes and momenta instead of only between coordinates.\\

On the other hand, the Lorentz algebra's anomaly term (\ref{99}) is simplified to:
\begin{equation}
	\label{lo}
	\begin{array}{*{20}{l}}
		{{T^{\mu \nu \rho \lambda }} = i{\pi ^2}\left( {\begin{array}{*{20}{l}}
					{\gamma _0^{\nu \lambda }{x^\mu }{x^\rho } + \gamma _0^{\nu \rho }{x^\mu }{x^\lambda }}\\
					{ + \gamma _0^{\mu \lambda }{x^\nu }{x^\rho } + \gamma _0^{\mu \rho }{x^\nu }{x^\lambda }}
			\end{array}} \right) + 2i{\pi ^2}\alpha '\tau \left( {\begin{array}{*{20}{l}}
					{\begin{array}{*{20}{l}}
							{\gamma _0^{\nu \rho }{x^\mu }{p^\lambda } - \gamma _0^{\mu \lambda }{x^\rho }{p^\nu }}\\
							{ + \gamma _0^{\nu \lambda }{x^\mu }{p^\rho } - \gamma _0^{\mu \rho }{x^\lambda }{p^\nu }}
					\end{array}}\\
					{\begin{array}{*{20}{l}}
							{ + \gamma _0^{\mu \rho }{x^\nu }{p^\lambda } - \gamma _0^{\nu \lambda }{x^\rho }{p^\mu }}\\
							{ + \gamma _0^{\mu \lambda }{x^\nu }{p^\rho } - \gamma _0^{\nu \rho }{x^\lambda }{p^\mu }}
					\end{array}}
			\end{array}} \right)}\\
		{ + \left( {i\theta _0^{\mu \rho } - 4i{\pi ^2}{{\alpha '}^2}{\tau ^2}\gamma _0^{\mu \rho }} \right){p^\lambda }{p^\nu } + \left( {i\theta _0^{\mu \lambda } - 4i{\pi ^2}{{\alpha '}^2}{\tau ^2}\gamma _0^{\mu \lambda }} \right){p^\rho }{p^\nu }}\\
		{\begin{array}{*{20}{l}}
				{ + \left( {i\theta _0^{\nu \rho } - 4i{\pi ^2}{{\alpha '}^2}{\tau ^2}\gamma _0^{\nu \rho }} \right){p^\lambda }{p^\mu } + \left( {i\theta _0^{\nu \lambda } - 4i{\pi ^2}{{\alpha '}^2}{\tau ^2}\gamma _0^{\nu \lambda }} \right){p^\rho }{p^\mu }}\\
				{}
		\end{array}}\\
	\end{array}
\end{equation}
\begin{equation}
	\label{jj}
	\begin{array}{l}
		{K^{\nu \mu \rho }} = 2i{\pi ^2}\alpha '\tau \gamma _0^{\nu \mu }{p^\rho } - 2i{\pi ^2}\alpha '\tau \gamma _0^{\rho \mu }{p^\nu }\\{\rm{~~~~~~~~}}
		+ i{\pi ^2}\gamma _0^{\mu \rho }{p^\nu } - i{\pi ^2}\gamma _0^{\mu \nu }{p^\rho }
	\end{array}
\end{equation}
Although the imposed condition on the deformation parameters (\ref{455}) eliminates all anomaly terms in the modified super-Virasoro algebra, the Lorentz algebra still contains a residual anomaly term (\ref{lo}) and (\ref{jj}). Therefore, as it is the case in several studies \cite{witten1, SSZABO, Carroll_2001,RevModPhys.73.977}, in the present work, Lorentz symmetry cannot be restored pertubatively at a first order.\\

\section {Summary and Results}
In this work, we have explored the effects of non-commutative phase-space on free open fermionic string theory. By introducing non-commutativity in both coordinates and momenta, we derived the modified super-Virasoro algebra for both the Ramond and Neveu-Schwarz sectors. This modification introduced additional anomaly terms, which could affect the consistency of the theory, including conformal invariance and the structure of the mass spectrum.\\
To resolve these issues, we imposed a specific relation between the non-commutativity parameters of space $\theta_{(m)}^{\mu\nu}$ and momentum $\gamma_{(m)}^{\mu\nu}$ (\ref{455}). Physically, this relation shows that the deformations of coordinates and momenta are not independent but part of a single consistent phase-space structure. It links the spatial and momentum noncommutativity in such a way that conformal symmetry is preserved (See Annexe \ref{ANNC}). In other words, the relation (\ref{455}) defines the exact balance required for noncommutative effects to exist without breaking the Virasoro symmetry. This condition, also, led to the cancellation of all Virasoro anomalies, allowing the algebra to recover its standard form while maintaining the presence of non-commutativity at the fundamental level. Additionally, a redefinition of the Fock space was necessary to diagonalize the mass operator and preserve the usual mass spectrum.\\
Notice that, it is the simultaneous presence of the non-commutative parameters $\theta$ and $\gamma$ that allows us to impose the necessary restrictions, demonstrating that non-commutative deformations can be incorporated into string theory preserving Virasoro algebra, provided that specific constraints are applied. This is the main motivation that led us to consider not a non-commutative spacetime, but a non-commutative phase-space.\\
{Notice that in the previous study \cite{Chaichian_2001,SSZABO}, non-commutativity in the target space was found to produce partial supersymmetry breaking. In the present approach the deformation acts on the phase-space and not directly on the target-space geometry. As a result, no explicit supersymmetry breaking is observed at first order. We restrict our conclusions to this perturbative regime.}\\
{Also, the analysis presented in this work is restricted to first order in the deformation parameters. Beyond this order, higher powers of the deformation parameters generate increasingly non-local contributions to the Virasoro generators.
	These terms might, aswell reintroduce new anomalies in the Virasoro and Lorentz algebra. A systematic investigation of these effects is left for future work.\\}
\section{Acknowledgment}
This work is supported by the Algerian Ministry of High Education and Research under the PRFU project No. B00L02UN250120220011.
\clearpage
\appendix
\section{Calculation of $[M^{\mu \nu },M^{\rho \lambda }]$} \label{222}
Using the equations (\ref{66}) and (\ref{6}) and (\ref{669}) one can find that:
\begin{equation}
	\begin{array}{l}
		\left[ {{x^\mu }{p^\nu },{x^\rho }{p^\lambda }} \right] = i{\pi ^2}\gamma _0^{\nu \lambda }{x^\mu }{x^\rho } 
		+ \left( {2i{\pi ^2}\alpha '\tau \gamma _0^{\nu \rho } - i{\eta ^{\nu \rho }}} \right){x^\mu }{p^\lambda }\\
		+ \left( {i{\eta ^{\mu \lambda }} - 2i{\pi ^2}\alpha '\tau \gamma _0^{\mu \lambda }} \right){x^\rho }{p^\nu } 
		+ \left( {i\theta _0^{\mu \rho } - 4i{\pi ^2}{{\alpha '}^2}{\tau ^2}\gamma _0^{\mu \rho }} \right){p^\lambda }{p^\nu }
	\end{array}
\end{equation}
\begin{equation}
	\begin{array}{l}
		\left[ {{x^\mu }{p^\nu },{x^\rho }{p^\lambda }} \right] = i{\pi ^2}\gamma _0^{\nu \lambda }{x^\mu }{x^\rho } 
		+\left( {2i{\pi ^2}\alpha '\tau \gamma _0^{\nu \rho } - i{\eta ^{\nu \rho }}} \right){x^\mu }{p^\lambda }\\
		+ \left( {i{\eta ^{\mu \lambda }} - 2i{\pi ^2}\alpha '\tau \gamma _0^{\mu \lambda }} \right){x^\rho }{p^\nu }
		+ \left( {i\theta _0^{\mu \rho } - 4i{\pi ^2}{{\alpha '}^2}{\tau ^2}\gamma _0^{\mu \rho }} \right){p^\lambda }{p^\nu }
	\end{array}
\end{equation}
\begin{equation}
	\begin{array}{l}
		\left[ {{x^\nu }{p^\mu },{x^\lambda }{p^\rho }} \right] = i{\pi ^2}\gamma _0^{\mu \rho }{x^\nu }{x^\lambda } + \left( {2i{\pi ^2}\alpha '\tau \gamma _0^{\mu \lambda } - i{\eta ^{\mu \lambda }}} \right){x^\nu }{p^\rho }\\
		+ \left( {i{\eta ^{\nu \rho }} - 2i{\pi ^2}\alpha '\tau \gamma _0^{\nu \rho }} \right){x^\lambda }{p^\mu } + \left( {i\theta _0^{\nu \lambda } - 4i{\pi ^2}{{\alpha '}^2}{\tau ^2}\gamma _0^{\nu \lambda }} \right){p^\rho }{p^\mu }
	\end{array}
\end{equation}
\begin{equation}
	\begin{array}{*{20}{l}}
		{\left[ {{x^\nu }{p^\mu },{x^\rho }{p^\lambda }} \right] = i{\pi ^2}\gamma _0^{\mu \lambda }{x^\nu }{x^\rho } + \left( {2i{\pi ^2}\alpha '\tau \gamma _0^{\mu \rho } - i{\eta ^{\mu \rho }}} \right){x^\nu }{p^\lambda }}\\
		{ + \left( {i{\eta ^{\nu \lambda }} - 2i{\pi ^2}\alpha '\tau \gamma _0^{\nu \lambda }} \right){x^\rho }{p^\mu } + \left( {i\theta _0^{\nu \rho } - 4i{\pi ^2}{{\alpha '}^2}{\tau ^2}\gamma _0^{\nu \rho }} \right){p^\lambda }{p^\mu }}\\
	\end{array}
\end{equation}
And for the mode part, we find that:\\
\begin{equation}
	\begin{array}{*{20}{l}}
		{\left[ {\alpha _{ - n}^\mu \alpha _n^\nu ,\alpha _{ - m}^\rho \alpha _m^\lambda } \right] = \left( \begin{array}{l}
				n{\eta ^{\nu \rho }} + i\frac{{{{(2\pi \alpha ')}^2}}}{{2\alpha '}}{\gamma _n}^{\nu \rho }\\
				+ i\frac{{{n^2}}}{{2\alpha '}}{\theta _n}^{\nu \rho }
			\end{array} \right)\alpha _{ - n}^\mu \alpha _n^\lambda }\\
		{ + \left( \begin{array}{l}
				- n{\eta ^{\mu \lambda }} + i\frac{{{{(2\pi \alpha ')}^2}}}{{2\alpha '}}{\gamma _n}^{\mu \lambda }\\
				+ i\frac{{{n^2}}}{{2\alpha '}}{\theta _n}^{\mu \lambda }
			\end{array} \right)\alpha _{ - n}^\rho \alpha _n^\nu }
	\end{array}
\end{equation}
\begin{equation}
	\begin{array}{*{20}{l}}
		{\left[ {\alpha _{ - n}^\mu \alpha _n^\nu ,\alpha _{ - m}^\lambda \alpha _m^\rho } \right] = \left( \begin{array}{l}
				n{\eta ^{\nu \lambda }} + i\frac{{{{(2\pi \alpha ')}^2}}}{{2\alpha '}}{\gamma _n}^{\nu \lambda }\\
				+ i\frac{{{n^2}}}{{2\alpha '}}{\theta _n}^{\nu \lambda }
			\end{array} \right)\alpha _{ - n}^\mu \alpha _n^\rho }\\
		{ + \left( \begin{array}{l}
				- n{\eta ^{\mu \rho }} + i\frac{{{{(2\pi \alpha ')}^2}}}{{2\alpha '}}{\gamma _n}^{\mu \rho }\\
				+ i\frac{{{n^2}}}{{2\alpha '}}{\theta _n}^{\mu \rho }
			\end{array} \right)\alpha _{ - n}^\lambda \alpha _n^\nu }
	\end{array}
\end{equation}
\begin{equation}
	\begin{array}{*{20}{l}}
		{\left[ {\alpha _{ - n}^\nu \alpha _n^\mu ,\alpha _{ - m}^\lambda \alpha _m^\rho } \right] = \left( \begin{array}{l}
				n{\eta ^{\mu \lambda }} + i\frac{{{{(2\pi \alpha ')}^2}}}{{2\alpha '}}{\gamma _n}^{\mu \lambda }\\
				+ i\frac{{{n^2}}}{{2\alpha '}}{\theta _n}^{\mu \lambda }
			\end{array} \right)\alpha _{ - n}^\nu \alpha _n^\rho }\\
		{ + \left( \begin{array}{l}
				- n{\eta ^{\nu \rho }} + i\frac{{{{(2\pi \alpha ')}^2}}}{{2\alpha '}}{\gamma _n}^{\nu \rho }\\
				+ i\frac{{{n^2}}}{{2\alpha '}}{\theta _n}^{\nu \rho }
			\end{array} \right)\alpha _{ - n}^\lambda \alpha _n^\mu }
	\end{array}
\end{equation}
\begin{equation}
	\begin{array}{*{20}{l}}
		{\left[ {\alpha _{ - n}^\nu \alpha _n^\mu ,\alpha _{ - m}^\rho \alpha _m^\lambda } \right] = \left( \begin{array}{l}
				n{\eta ^{\mu \rho }} + i\frac{{{{(2\pi \alpha ')}^2}}}{{2\alpha '}}{\gamma _n}^{\mu \rho }\\
				+ i\frac{{{n^2}}}{{2\alpha '}}{\theta _n}^{\mu \rho }
			\end{array} \right)\alpha _{ - n}^\nu \alpha _n^\lambda }\\
		{ + \left( \begin{array}{l}
				- n{\eta ^{\nu \lambda }} + i\frac{{{{(2\pi \alpha ')}^2}}}{{2\alpha '}}{\gamma _n}^{\nu \lambda }\\
				+ i\frac{{{n^2}}}{{2\alpha '}}{\theta _n}^{\nu \lambda }
			\end{array} \right)\alpha _{ - n}^\rho \alpha _n^\mu }
	\end{array}
\end{equation}
\begin{equation}
	\begin{array}{*{20}{l}}
		{\begin{array}{*{20}{l}}
				{\left[ {b_{ - r}^\mu b_r^\nu ,b_{ - s}^\rho b_s^\lambda } \right] = 2b_{ - r}^\mu b_{ - s}^\rho b_r^\nu b_s^\lambda  + 2b_{ - r}^\mu b_r^\nu b_{ - s}^\rho b_s^\lambda }\\
				{ + 2b_{ - s}^\rho b_{ - r}^\mu b_s^\lambda b_r^\nu  + 2b_{ - r}^\mu b_{ - s}^\rho b_s^\lambda b_r^\nu  - {\eta ^{\nu \lambda }}b_{ - r}^\mu b_r^\rho }
		\end{array}}\\
		{\begin{array}{*{20}{l}}
				{ - {\eta ^{\nu \rho }}b_{ - r}^\mu b_r^\lambda  - {\eta ^{\mu \lambda }}b_{ - r}^\rho b_r^\nu  - {\eta ^{\mu \rho }}b_{ - r}^\lambda b_r^\nu }\\				
		\end{array}}
	\end{array}
\end{equation}
\begin{equation}
	\begin{array}{*{20}{l}}
		{\begin{array}{*{20}{l}}
				{\left[ {b_{ - r}^\nu b_r^\mu ,b_{ - s}^\rho b_s^\lambda } \right] = 2b_{ - r}^\nu b_{ - s}^\rho b_r^\mu b_s^\lambda  + 2b_{ - r}^\nu b_r^\mu b_{ - s}^\rho b_s^\lambda }\\
				{ + 2b_{ - s}^\rho b_{ - r}^\nu b_s^\lambda b_r^\mu  + 2b_{ - r}^\nu b_{ - s}^\rho b_s^\lambda b_r^\mu  - {\eta ^{\mu \lambda }}b_{ - r}^\nu b_r^\rho }
		\end{array}}\\
		{\begin{array}{*{20}{l}}
				{ - {\eta ^{\mu \rho }}b_{ - r}^\nu b_r^\lambda  - {\eta ^{\nu \lambda }}b_{ - r}^\rho b_r^\mu  - {\eta ^{\nu \rho }}b_{ - r}^\lambda b_r^\mu }\\
		\end{array}}
	\end{array}
\end{equation}
\begin{equation}
	\begin{array}{*{20}{l}}
		{\begin{array}{*{20}{l}}
				{\left[ {b_{ - r}^\mu b_r^\nu ,b_{ - s}^\rho b_s^\lambda } \right] = 2b_{ - r}^\mu b_{ - s}^\lambda b_r^\nu b_s^\rho  + 2b_{ - r}^\mu b_r^\nu b_{ - s}^\lambda b_s^\rho }\\
				{ + 2b_{ - s}^\lambda b_{ - r}^\mu b_s^\rho b_r^\nu  + 2b_{ - r}^\mu b_{ - s}^\lambda b_s^\rho b_r^\nu  - {\eta ^{\nu \rho }}b_{ - r}^\mu b_r^\lambda }
		\end{array}}\\
		{\begin{array}{*{20}{l}}
				{ - {\eta ^{\nu \lambda }}b_{ - r}^\mu b_r^\rho  - {\eta ^{\mu \rho }}b_{ - r}^\lambda b_r^\nu  - {\eta ^{\mu \lambda }}b_{ - r}^\rho b_r^\nu }\\
		\end{array}}
	\end{array}
\end{equation}
\begin{equation}
	\begin{array}{*{20}{l}}
		{\begin{array}{*{20}{l}}
				{\left[ {b_{ - r}^\nu b_r^\mu ,b_{ - s}^\lambda b_s^\rho } \right] = 2b_{ - r}^\nu b_{ - s}^\lambda b_r^\mu b_s^\rho  + 2b_{ - r}^\nu b_r^\mu b_{ - s}^\lambda b_s^\rho }\\
				{ + 2b_{ - s}^\lambda b_{ - r}^\nu b_s^\rho b_r^\mu  + 2b_{ - r}^\nu b_{ - s}^\lambda b_s^\rho b_r^\mu  - {\eta ^{\mu \rho }}b_{ - r}^\nu b_r^\lambda }
		\end{array}}\\
		{\begin{array}{*{20}{l}}
				{ - {\eta ^{\mu \lambda }}b_{ - r}^\nu b_r^\rho  - {\eta ^{\nu \rho }}b_{ - r}^\lambda b_r^\mu  - {\eta ^{\nu \lambda }}b_{ - r}^\rho b_r^\mu }\\
		\end{array}}
	\end{array}
\end{equation}
Now, we use the equation (\ref{98}) to calculate $[{M^{\mu \nu }},{M^{\rho \lambda }}]$ :
\begin{equation}
	\begin{array}{l} [{M^{\mu \nu }},{M^{\rho \lambda }}] =  - i{\eta ^{\nu \rho }}{M^{\mu \lambda }} + i{\eta ^{\mu \lambda }}{M^{\rho \nu }} + i{\eta ^{\nu \lambda }}{M^{\mu \rho }}  - i{\eta ^{\mu \rho }}{M^{\lambda \nu }} +i{\pi ^2}\left( \begin{array}{l} \gamma _0^{\nu \lambda }{x^\mu }{x^\rho } + \gamma _0^{\nu \rho }{x^\mu }{x^\lambda } +\\ \gamma _0^{\mu \lambda }{x^\nu }{x^\rho } + \gamma _0^{\mu \rho }{x^\nu }{x^\lambda } \end{array} \right)\\
		+ 2i{\pi ^2}\alpha '\tau \left( \begin{array}{l}
			\gamma _0^{\nu \rho }{x^\mu }{p^\lambda } - \gamma _0^{\mu \lambda }{x^\rho }{p^\nu } \\
			+ \gamma _0^{\nu \lambda }{x^\mu }{p^\rho } - \gamma _0^{\mu \rho }{x^\lambda }{p^\nu } + \\
			\gamma _0^{\mu \rho }{x^\nu }{p^\lambda } - \gamma _0^{\nu \lambda }{x^\rho }{p^\mu } \\
			+ \gamma _0^{\mu \lambda }{x^\nu }{p^\rho } - \gamma _0^{\nu \rho }{x^\lambda }{p^\mu }
		\end{array} \right) + 
		\left( {i\theta _0^{\mu \rho } - 4i{\pi ^2}{{\alpha '}^2}{\tau ^2}\gamma _0^{\mu \rho }} \right){p^\lambda }{p^\nu } 
		+ \left( {i\theta _0^{\mu \lambda } - 4i{\pi ^2}{{\alpha '}^2}{\tau ^2}\gamma _0^{\mu \lambda }} \right){p^\rho }{p^\nu } +\\
		\left( {i\theta _0^{\nu \rho } - 4i{\pi ^2}{{\alpha '}^2}{\tau ^2}\gamma _0^{\nu \rho }} \right){p^\lambda }{p^\mu } 
		+ \left( {i\theta _0^{\nu \lambda } - 4i{\pi ^2}{{\alpha '}^2}{\tau ^2}\gamma _0^{\nu \lambda }} \right){p^\rho }{p^\mu } + 
		\left( {i\frac{{{{(2\pi \alpha ')}^2}}}{{2\alpha '}}{\gamma _n}^{\nu \rho } + i\frac{{{n^2}}}{{2\alpha '}}{\theta _n}^{\nu \rho }} \right)\left( {\alpha _{ - n}^\mu \alpha _n^\lambda  + \alpha _{ - n}^\lambda \alpha _n^\mu } \right)\\ + 
		\left( {i\frac{{{{(2\pi \alpha ')}^2}}}{{2\alpha '}}{\gamma _n}^{\mu \lambda } + i\frac{{{n^2}}}{{2\alpha '}}{\theta _n}^{\mu \lambda }} \right)\left( {\alpha _{ - n}^\rho \alpha _n^\lambda  + \alpha _{ - n}^\lambda \alpha _n^\rho } \right) + 
		\left( {i\frac{{{{(2\pi \alpha ')}^2}}}{{2\alpha '}}{\gamma _n}^{\nu \lambda } + i\frac{{{n^2}}}{{2\alpha '}}{\theta _n}^{\nu \lambda }} \right)\left( {\alpha _{ - n}^\rho \alpha _n^\mu  + \alpha _{ - n}^\mu \alpha _n^\rho } \right) + \\
		\left( {i\frac{{{{(2\pi \alpha ')}^2}}}{{2\alpha '}}{\gamma _n}^{\mu \rho } + i\frac{{{n^2}}}{{2\alpha '}}{\theta _n}^{\mu \rho }} \right)\left( {\alpha _{ - n}^\lambda \alpha _n^\nu  + \alpha _{ - n}^\nu \alpha _n^\lambda } \right)
	\end{array}
\end{equation}

%-----------------------------------------------------------------------------------

\section{Derivation of the Equations of Motion from the Deformed Action}
\label{appb}
Starting from the deformed world-sheet action:
\begin{equation}
	S_* = -\frac{1}{4\pi\alpha'} 
	\int d^2\sigma
	\Big(
	\partial_a X^\mu * \partial^a X_\mu
	- i\,\bar{\psi}^\mu * \rho^a \partial_a \psi_\mu
	\Big),
	\label{B1}
\end{equation}
we compute the functional variations with respect to the fields $X^\mu$ and $\bar{\psi}^\mu$.

The noncommutative structure of the world-sheet is defined by:
\begin{equation}
	[\sigma^a, \sigma^b] = i\,\theta^{ab}, \qquad
	\theta^{ab} = -\theta^{ba},
\end{equation}
and the Moyal (star) product of two fields $A$ and $B$ is given by
\begin{equation}
	A*B = A\,e^{\frac{i}{2}\theta^{cd}\overleftarrow{\partial_c}\overrightarrow{\partial_d}} B
	= AB + \frac{i}{2}\theta^{cd}(\partial_c A)(\partial_d B) + \mathcal{O}(\theta^2).
	\label{B2}
\end{equation}
We retain only first-order terms in $\theta^{ab}$ throughout.\\
Applying \eqref{B2} to the bosonic and fermionic parts of \eqref{B1} gives:

\paragraph{Bosonic term:}
\begin{equation}
	\partial_a X^\mu * \partial^a X_\mu
	= (\partial_a X^\mu)(\partial^a X_\mu)
	+ \frac{i}{2}\theta^{cd}(\partial_c\partial_a X^\mu)(\partial_d\partial^a X_\mu).
	\label{B3}
\end{equation}

\paragraph{Fermionic term:}
\begin{equation}
	\bar{\psi}^\mu * \rho^a \partial_a \psi_\mu
	= \bar{\psi}^\mu\rho^a\partial_a\psi_\mu
	+ \frac{i}{2}\theta^{cd}(\partial_c\bar{\psi}^\mu)\rho^a(\partial_d\partial_a\psi_\mu).
	\label{B4}
\end{equation}

Substituting \eqref{B3} and \eqref{B4} into \eqref{B1} gives the expanded action:
\begin{equation}
	\label{B5} {S_*} =  - \frac{1}{{4\pi \alpha '}}\int {{d^2}} \sigma \left[ \begin{array}{l}
		({\partial _a}{X^\mu })({\partial ^a}{X_\mu }) + 
		\frac{i}{2}{\theta ^{cd}}({\partial _c}{\partial _a}{X^\mu })({\partial _d}{\partial ^a}{X_\mu }) - \\
		i{\mkern 1mu} {{\bar \psi }^\mu }{\rho ^a}{\partial _a}{\psi _\mu } + 
		\frac{1}{2}{\theta ^{cd}}({\partial _c}{{\bar \psi }^\mu }){\rho ^a}({\partial _d}{\partial _a}{\psi _\mu })
	\end{array} \right]
\end{equation}

We separate the commutative and first-order terms:
\begin{align}
	S_*^{(X)} = S_0^{(X)} + S_1^{(X)},
\end{align}
where
\begin{equation}
	\label{B6} \begin{array}{l}
		S_0^{(X)} =  - \frac{1}{{4\pi \alpha '}}\int {{d^2}} \sigma {\mkern 1mu} ({\partial _a}{X^\mu })({\partial ^a}{X_\mu }),\\
		S_1^{(X)} =  - \frac{i}{{8\pi \alpha '}}\int {{d^2}} \sigma {\mkern 1mu} {\theta ^{cd}}({\partial _c}{\partial _a}{X^\mu })({\partial _d}{\partial ^a}{X_\mu })
	\end{array}	
\end{equation}
From the ordinary part, one get the ordinary solution, so that:
\begin{equation}
	\delta S_0^{(X)} 
	= +\frac{1}{2\pi\alpha'}\int d^2\sigma\,
	(\partial_a\partial^a X_\mu)\,\delta X^\mu
	\label{B8}
\end{equation}
and $S_1^{(X)}$ do not contribute in the equations of motion, indeed:
\begin{align}
	\begin{array}{*{20}{l}}
		{\delta S_1^{(X)}}&{ =  - \frac{i}{{8\pi \alpha '}}\int  {d^2}\sigma {\mkern 1mu} {\theta ^{cd}}\left[ \begin{array}{l}
				({\partial _c}{\partial _a}\delta {X^\mu })({\partial _d}{\partial ^a}{X_\mu }) + 
				({\partial _c}{\partial _a}{X^\mu })({\partial _d}{\partial ^a}\delta {X_\mu })
			\end{array} \right]}
	\end{array}
\end{align}
Since both terms are equal, we write:
\begin{equation}
	\delta S_1^{(X)}
	= -\frac{i}{4\pi\alpha'}\int d^2\sigma\,
	\theta^{cd}(\partial_c\partial_a X_\mu)(\partial_d\partial^a \delta X^\mu).
	\label{B9}
\end{equation}
Integrating by parts first with respect to $\partial_d$ and then with respect to $\partial_a$,
and neglecting boundary terms, gives:
\begin{equation}
	\delta S_1^{(X)}
	= -\frac{i}{4\pi\alpha'}\int d^2\sigma\,
	\theta^{cd}(\partial_a\partial_d\partial_c\partial^a X_\mu)\,\delta X^\mu
	\label{B10}
\end{equation}

Now, combining \eqref{B8} and \eqref{B10}:
\begin{equation}
	\delta S_*^{(X)}
	= \frac{1}{2\pi\alpha'}\int d^2\sigma\,
	\Big[\partial_a\partial^a X_\mu
	- \frac{i}{2}\theta^{cd}\partial_a\partial_d\partial_c\partial^a X_\mu\Big]
	\delta X^\mu.
	\label{B11}
\end{equation}
Since $\theta^{cd}$ is antisymmetric while $\partial_d\partial_c$ is symmetric, the second term vanishes. 
The Euler--Lagrange equation therefore reduces to:
\begin{equation}
	(\partial_\tau^2 - \partial_\sigma^2)X^\mu = 0
	\label{B13}
\end{equation}

Now, Variation with respect to {$\bar{\psi}^\mu$}, from \eqref{B5}, the fermionic Lagrangian density is:
\begin{equation}
	\mathcal{L}_\psi
	=\frac{i}{4\pi\alpha'}
	\Big[
	\bar{\psi}^\mu\rho^a\partial_a\psi_\mu
	-\frac{i}{2}\theta^{cd}(\partial_c\bar{\psi}^\mu)\rho^a(\partial_d\partial_a\psi_\mu)
	\Big].
	\label{B14}
\end{equation}
same calculations leading to the standard Dirac-type equation:
\begin{equation}
	\rho^a\partial_a\psi^\mu = 0
	\label{B16}
\end{equation}

\section{Conformal Invariance Verification}\label{ANNC}
We verify explicitly that the Virasoro generators defined in \eqref{11}, \eqref{12}, \eqref{13} and \eqref{14} continue to generate conformal transformations of the fields $X^\mu$ and $\psi^\mu$
in the presence of the deformed commutation relations \eqref{66}, \eqref{6} and \eqref{669},
provided the noncommutative structure tensors satisfy the consistency condition:
\begin{equation}
	\gamma_{(m)}^{\mu\nu} = -\frac{m^2}{\left(2\pi\alpha'\right)^2}\,\theta_{(m)}^{\mu\nu}
\end{equation}

\vspace{6pt}
\noindent\textbf{Bosonic field:}

First we Calclate the commutator between $L_m$ and $\alpha_n^\mu$. From \eqref{13},
\begin{equation}
	L_m = \frac{1}{2}\sum_k :\!\alpha_{m-k}^\rho \alpha_{k\,\rho}\!:
	+ \frac{1}{4}\sum_r (2r - m):\!b_{m-r}^\rho b_{r\,\rho}\!:,
\end{equation}
and using the deformed oscillator algebra \eqref{6},
\begin{equation}
	[\alpha_m^\mu, \alpha_n^\nu]
	= \left(m\eta^{\mu\nu}
	+ i\frac{(2\pi\alpha')^2}{2\alpha'}\,\gamma_n^{\mu\nu}
	+ i\frac{n^2}{2\alpha'}\,\theta_n^{\mu\nu}\right)\delta_{m+n,0}.
	\label{eq:B6_comm}
\end{equation}
Computing the commutator:
\begin{align}
	[L_m, \alpha_n^\mu]
	&= \frac{1}{2}\sum_k
	\big(
	[\alpha_{m-k}^\rho, \alpha_n^\mu]\alpha_{k\,\rho}
	+ \alpha_{m-k}^\rho[\alpha_{k\,\rho}, \alpha_n^\mu]
	\big).
\end{align}
Substituting \ref{eq:B6_comm} and simplifying using the delta functions gives:
\begin{equation}
	[L_m, \alpha_n^\mu]
	= -\,n\,\alpha_{m+n}^\mu
	+ i\frac{(2\pi\alpha')^2}{4\alpha'}\,\gamma_n^{\mu}{}_{\rho}\alpha_{m+n}^\rho
	+ i\frac{n^2}{4\alpha'}\,\theta_n^{\mu}{}_{\rho}\alpha_{m+n}^\rho
	\label{eq:B6_comm_alpha}
\end{equation}
where the first term reproduces the standard conformal transformation;
the remaining terms are the noncommutative corrections.\\

Second, we calculate the action of $L_m$ on $X^\mu(\sigma,\tau)$, from \eqref{4}:
\begin{equation}
	X^\mu(\sigma,\tau)
	= x^\mu + 2\alpha' p^\mu\tau
	+ i\sqrt{2\alpha'}\sum_{n\neq0}\frac{1}{n}\alpha_n^\mu e^{-in\tau}\cos n\sigma.
\end{equation}
applying \ref{eq:B6_comm_alpha},
\begin{equation}
	\begin{array}{l}
		[{L_m},{X^\mu }] = i\sqrt {2\alpha '} \sum\limits_{n \ne 0} {\frac{1}{n}} [{L_m},\alpha _n^\mu ]{e^{ - in\tau }}\cos n\sigma \\
		=  - {\mkern 1mu} i\sqrt {2\alpha '} \sum\limits_{n \ne 0} {\alpha _{m + n}^\mu } {e^{ - in\tau }}\cos n\sigma  + \\
		\frac{{i\sqrt {2\alpha '} }}{{4\alpha '}}\sum\limits_{n \ne 0} {\frac{1}{n}} \left[ \begin{array}{l}
			{(2\pi \alpha ')^2}\gamma_{n\rho}^{\mu} + \\
			{n^2}\theta_{n\rho}^{\mu} 
		\end{array} \right]\alpha _{m + n}^\rho {e^{ - in\tau }}\cos n\sigma 
	\end{array}
	\label{eq:B6_action}
\end{equation}	
By Shifting the summation index $n\!\rightarrow\! n-m$ in the first term of \eqref{eq:B6_action}
and use trigonometric identities to identify the derivative structure
of $\partial_+ X^\mu = (\partial_\tau + \partial_\sigma)X^\mu$.
This yields:
\begin{equation}
	\begin{array}{l}
		[{L_m},{X^\mu }] = i{\mkern 1mu} {e^{im(\tau  + \sigma )}}{\partial _ + }{X^\mu } + \\
		\frac{i}{{4\alpha '}}\left[ {{{(2\pi \alpha ')}^2}\gamma {{_n^\mu }_\rho } + {n^2}\theta {{_n^\mu }_\rho }} \right]{e^{im(\tau  + \sigma )}}{\partial _ + }{X^\rho }.
	\end{array}
	\label{eq:B6_action_pre}
\end{equation}

Applying the structural relation between $\gamma_{(m)}$ and $\theta_{(m)}$ \eqref{456} and substituting  into (\ref{eq:B6_action_pre}) yields:
\begin{equation}
	[L_m, X^\mu(\sigma,\tau)] = i\,e^{im(\tau+\sigma)}\,\partial_+ X^\mu(\sigma,\tau)
	\label{eq:B6_final_X}
\end{equation}
Hence, the Virasoro generators act on $X^\mu$ exactly as in the commutative theory.

\vspace{6pt}
\noindent\textbf{Fermionic field:}

The field $\psi^\mu $ is not affected by noncommutativity, and the reasoning follows the same logic as in the ordinary case. Using \eqref{669} and the fermionic part of $L_m$,
\begin{equation}
	[L_m, b_s^\mu] = \left(\frac{m}{2}-s\right)b_{m+s}^\mu,
\end{equation}
and therefore, in coordinate space,
\begin{equation}
	[L_m, \psi^\mu(\sigma,\tau)] =
	i\,e^{im(\tau+\sigma)}\!\left(\partial_+ + \frac{i m}{2}\right)\psi^\mu(\sigma,\tau)
	\label{eq:B6_final_psi}
\end{equation}

Both bosonic and fermionic sectors preserve their standard conformal transformation
laws, demonstrating that conformal invariance holds exactly in the deformed theory.

\end{document}